# On the Prediction of Hardware Security Properties of HLS Designs Using Graph Neural Networks

Amalia-Artemis Koufopoulou [1], Athanasios Papadimitriou [1] [2], Aggelos Pikrakis [1], Mihalis Psarakis [1] and David Hely[3]

[1] Dept. of Informatics, University of Piraeus, Piraeus, Greece
[2] Dept. of Digital Systems, University of the Peloponnese, Sparta, Greece
[3] Univ. Grenoble Alpes, F-26000 Valence, France
amaliakoufopulou@gmail.com, a.papadimitriou@uop.gr, mpsarak@unipi.gr, pikrakis@unipi.gr, david.hely@lcis.grenoble-inp.fr

*Abstract*—High-level synthesis (HLS) tools have provided significant productivity enhancements to the design flow of digital systems in recent years, resulting in highly-optimized circuits, in terms of area and latency. Given the evolution of hardware attacks, which can render them vulnerable, it is essential to consider security as a significant aspect of the HLS design flow. Yet the need to evaluate a huge number of functionally equivalent designs of the HLS design space challenges hardware security evaluation methods (e.g., fault injection - FI campaigns). In this work, we propose an evaluation methodology of hardware security properties of HLS-produced designs using state-of-the-art Graph Neural Network (GNN) approaches that achieves significant speedup and better scalability than typical evaluation methods (such as FI). We demonstrate the proposed methodology on a Double Modular Redundancy (DMR) countermeasure applied on an AES SBox implementation, enhanced by diversifying the redundant modules through HLS directives. The experimental results show that GNNs can be efficiently trained to predict important hardware security metrics concerning fault attacks (e.g., critical and detection error rates), by using regression. The proposed method predicts the fault vulnerability metrics of the HLS-based designs with high R-squared scores and achieves huge speedup compared to fault injection once the training of the GNN is completed.

## I. INTRODUCTION

Over the last decades, High-level Synthesis (HLS) has been proven a significant ally in developing quality designs targeting hardware platforms [1]. The methodology allows the coding of complex functionalities in widely accessible high-level languages and automatically generates their register transfer-level (RTL) equivalent. This revolutionizes the design flow in terms of productivity and democratizes the domain of developing hardware to non-experts [2]. Modern HLS tools offer optimization and design exploration capabilities within their flow, through compilation directives that can handle architectural considerations related to the design specifications [3]. The HLS paradigm role within the Electronic Design Automation (EDA) flow can be proved crucial, considering the increasing demand for acceleration, enabled by modern hardware platforms [2]. Thus, the advantages of HLS can significantly assist developers to respond to the strict time-to-market needs by implementing high-quality designs.

In many cases, hardware accelerators perform computations on sensitive data, as is the case of cryptographic [4] and Machine Learning [5] applications. Therefore, an additional important design aspect of these accelerators is their security level against attacks. Hardware attacks form a powerful threat against the security of these circuits and include fault and side-channel attacks [6]. For instance, the injection of faults into the operation of a hardware accelerator can severely affect its security and reliability, as studied in [7].

One way to increase the resilience of accelerators against such attacks is the integration of the appropriate countermeasures [8]. Though such countermeasures can be integrated at any level of abstraction (HLS, RTL, Gate, Layout), the most convenient and efficient level is the higher level used to describe the functionality of the accelerator. Lower abstraction levels might need to be used in case the countermeasure requires lower-level information, to be implemented efficiently [9], at the cost of additional complexity and effort. In [10] and [11], countermeasures were applied outside of the HLS flow to avoid any optimization that could reduce the expected levels of security. While such a methodology can assure that the HLS flow will not affect security properties, at the same time it loses all the advantages HLS offers.

Since HLS tools and optimizations do not consider hardware security properties, it becomes imperative to perform evaluations and search the HLS design space to find the optimal solutions concerning security [12][13]. Yet, traditional evaluation methods, such as fault injection (FI) campaigns [14] for evaluating fault detection countermeasures, require considerable time and computational efforts. Emulation and simulation techniques need to be employed to inject large numbers of faults, making these methods non-scalable, especially for complex circuits. As a compromise, statistical approaches are usually applied, at the cost of reduced accuracy. Statistical FI [15] inherently involves a margin of error and confidence level depending on the number of fault samples of the campaign.

In this work, we present a methodology based on state-of-the-art Graph Neural Networks (GNN) [16] proposing an accelerated approach to evaluate the HLS design space, i.e., all the functionally equivalent different implementations of the "protected" hardware accelerator that can be produced by the HLS design flow, against such attacks. We take advantage of the ability to represent any design's RTL netlist as a graph, so as to provide to a GNN the structure of the circuit. Once the GNN is trained to evaluate the efficiency of a countermeasure (e.g., the error detection coverage of a scheme that protects against fault attacks), it is able to predict error rate metrics in a small fraction of the time needed to perform a FI campaign. We study our proposed flow using an on-the-fly AES SBox high-level description [17] protected against fault attacks using double modular redundancy (DMR) [18]. In addition, we utilize the HLS synthesis directives to introduce a level of functional diversity [19], in an attempt to further enhance the fault detection capability of the DMR countermeasure against multiple (e.g., double)-fault attacks. The use of HLS synthesis directives to diversify the DMR replicas allowed us to generate a large number of functionally equivalent designs, assisting with the creation of a dataset for our model's training. The proposed approach can be integrated into the HLS design flow in order to evaluate the resilience of large quantities of HLS design space solutions. This way, it is possible to approximate the error detection capability of many HLS-generated diversified DMR schemes and, thus, to avoid the need to perform time-consuming FI campaigns. The experimental results showed that our proposed approach speeds up the evaluation by several orders of magnitude (depending on the total number of designs, with a small loss of accuracy).

The methodology and tools described in the current work will be published with an open-source license to the following link: https://github.com/*******

The paper is organized as follows. Section II presents the necessary background on the topics of GNNs and hardware fault tolerance, the efforts of introducing reliability in HLS flows and the concept of diversity as a fault tolerance countermeasure. Section III describes the methodology we followed to generate the designs under test. Section IV presents the design space under test. In Section V we present our results and lastly, in Section VI we provide our conclusions and potential future directions.

## II. RELATED WORK

### A. Graph Neural Networks for Security

Learning-based approaches have already been developed to leverage the information of graph-like structures. Previously explored methods include Recurrent Neural Networks (RNNs) [20] and Convolutional Neural Networks (CNNs), which are, however, limited to regular structures (e.g., images) [21]. Graph Neural Networks (GNNs) can be viewed as a generalization of CNNs, extending their use for irregular structures, for applications mainly related to social networks and biochemical components [22].

Recent approaches apply GNNs as a tool for assisting either the design or the evaluation of circuits [23]. Regarding the use of GNNs for circuit evaluation, in [24], the authors rely on them to perform reverse engineering by classifying subcircuits depending on their functionalities. In [25] and [26], the authors use GNNs as an alternative to FI-based reliability evaluation. They train GNNs by means of FIs to the same circuit so as to model fault tolerance metrics of individual flip-flops. Therefore, their flow uses GNNs to perform tasks characterizing the nodes of the graph. In the current work we present a graph-oriented approach, resulting in a global metric.

To the best of our knowledge, no work exists on GNN-based graph-oriented prediction of the security or reliability (e.g., error detection) metrics of countermeasures integrated at the HLS. The graphs used for training the GNNs originate from the circuit's RTL netlist, allowing faster evaluations (compared to gate-level evaluations) and analysis at the abstraction level following HLS. This way, the evaluation takes into account the effects of HLS on the integrated countermeasures, without the impact of the synthesis flow following RTL. So far, mainly gate-level graphs are used in literature to provide the GNNs with more accurate information at a lower level of abstraction.

### B. Countermeasures against fault attacks

Hardware redundancy is a well-established fault detection method [18] that replicates the protected component into several copies. Typically, these copies are executed in parallel, and comparison logic (either a comparator or a majority selection voter) is used to detect - and potentially correct - the errors. Hardware redundancy provides stronger protection against fault attacks than other, less resource intense redundancy techniques, such as temporal or information redundancy.

The concept of diversity is applied to enhance the error detection capabilities of the hardware redundancy techniques.Such countermeasures are designed to compute the same result in a different manner [27]. This way, the redundant modules cannot be easily affected in the same way. Diversity can be applied at the algorithmic level [19] by using different algorithms to produce the same results.

### C. Fault Tolerance in HLS

Attempts to introduce those countermeasures automatically in the context of an HLS flow have been proposed in the literature. In [10], the authors apply partial Triple Modular Redundancy (TMR) at the RTL output of HLS implementations. In order to determine the logic which needs to be triplicated, they use information from the HLS flow. Their approach leads to the minimization of the area overhead and also avoids the cross-optimizations between redundant modules. In [11], the authors apply hardware duplication to the control logic, which is considered a critical part of the circuit, by taking advantage of the access to the code's intermediate representation created at the beginning of the HLS flow.

Techniques to develop error-resilient circuits can also be directly integrated into the HLS flow. In [28], the authors treat fault tolerance as a design constraint and examine it, during the HLS flow, in parallel to the traditional constraints of area and latency. Works such as [29] explore the notion of a reliability-aware HLS design space, achieving better results than applying post-HLS TMR. In both cases, the authors quantify the fault tolerance by performing a theoretical analysis instead of actual FI campaigns, an approach that may lead to inaccurate results.

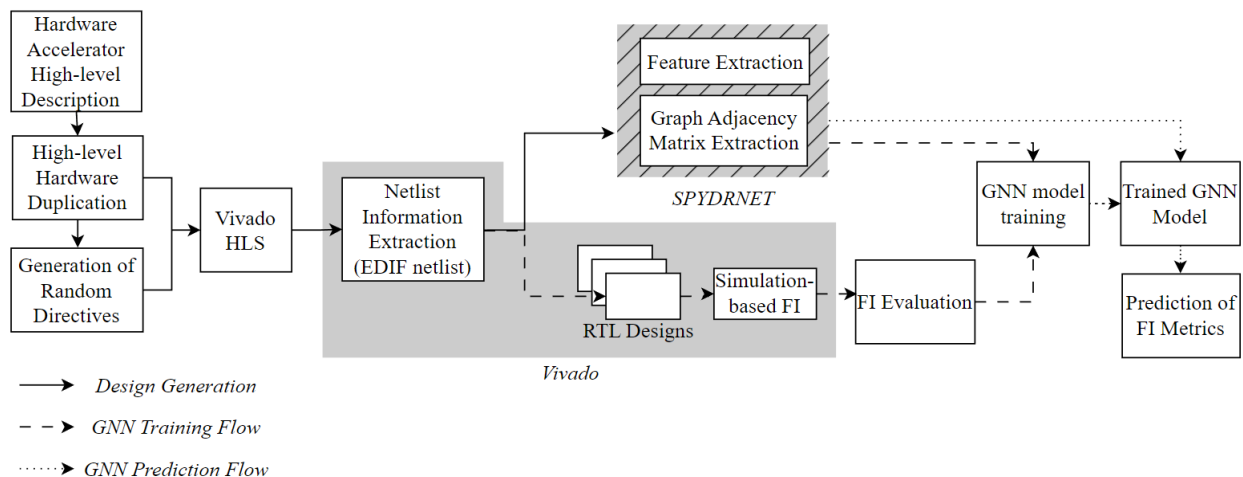

*Figure 1. GNN Training & Prediction Flow*

When combined with hardware redundancy, it is well-known that diversity significantly improves fault tolerance rates [30]. Thus, it is important to quantify diversity, especially in an early design stage, and possibly avoid costly FI simulations. To the best of our knowledge, the following three methods exist in literature. D-metric [30] examines the probability of double faults affecting the outputs of a diverse DMR implementation in the same way, so that the modules produce the same erroneous output. The methodology requires the exhaustive simulation of all possible faults (time and location of faults), for all possible inputs of the design under evaluation. Hence, even for relatively small designs, this approach is not scalable for the majority of existing fault models. Another metric presented in the literature is the Diversity Metric based on circuit Path analysis (DIMP) [31]. In this case, the metric seeks the same patterns of gates among computational paths of the redundant modules of a countermeasure, yet the authors only present its use against timing attacks. Both methodologies require the examination of the circuit either at RTL or gate-level, with the latter achieving higher accuracy. In [32], the authors develop a predictive model, trained with the HLS reports containing high-level characteristics (e.g., latency, number of FFs, LUTs, etc.) of multiple DMR designs, and their corresponding gate-level D-metric. After the completion of the training phase, the model is used to predict the diversity of a design. Their model presents better results compared to DIMP and RTL D-metric.

However, this approach may lead to misleading results. Specifically, in order to maximize diversity (i.e., by using a genetic algorithm), the methodology will attempt to differentiate as much as possible high-level properties (e.g., resources and timing/latency), resulting in sub-optimal designs, while better choices might exist in the HLS design space. Additionally, the approach in [32] is limited to a diversity-oriented countermeasure. Instead, our methodology can be generalized to examine any fault-tolerance countermeasure.

## III. EVALUATION FLOW

The evaluation methodology and tools we propose are depicted in Fig. 1. The flow is separated into the GNN training flow and the GNN prediction flow. The following sub-sections detail each of these flows.

### A. GNN Training Flow

Given a high-level description (in C language) of a duplicated hardware accelerator, we apply a random selection of Vivado HLS directives set to each redundant module in order to diversify them. A random synthesis directive is generated for each function and loop existing in the two modules.

*1) Generation of Random Directives and HLS Synthesis*

Firstly, the high-level input code is parsed to derive the names of the functions and loops included. Afterwards, a set of directives is applied, selected randomly from the collections of the function directives and the loop directives. Regarding functions, the collection includes the limitation of allocation in modules and/or operational units, imposing expression balancing, inlining and pipelining. For loops, we select the application of a dataflow directive, pipelining or unrolling. Each of these directives can be differentiated through various synthesis options; for example, unrolling can be performed by a different factor (in our flow, it can be set to 2, 3, 5 or 7). Our aim is to be able to choose from a generic and broad pool of designs, and thus, there may exist cases that a directive has no actual effect on the function/loop (i.e., a directive may impose the limitation of shifter logic on functions that do not contain any). In addition, the application of a directive in one replica does not prohibit the use of the same directive in another replica. In that case, the same directive could be applied in the corresponding functions/loops of two replicas and not diversify them. This would result in designs with different degrees of diversity.

Since Vivado HLS seeks optimization opportunities, the tool will enforce the two modules to operate sequentially, in order to perform resource sharing. Essentially, the countermeasure will serve as temporal rather than hardware redundancy, which was the initial goal. To remedy this behavior, an additional directive, #pragma HLS INTERFACE ap_ctrl_hs register port=return, was added to the top function of each DMR replica. The input code, along with the generated directive file, compose a solution of the implementation. We generate as many solutions as we need. Solutions are synthesized, producing the HDL functional equivalents. The HDL files are added to a Vivado project, from where the RTL netlists, as well as the registers for the FI evaluation, are extracted for each design.

*2) Netlist Information Extraction*

In order to extract the graph from each RTL netlist, as well useful information to annotate the graph elements, we use the SPYDRNET framework [33]. Using the EDIF netlist of each design as input, the framework allows us to extract the information that we consider relevant to the graph prediction task. GNN models require the use of the graph adjacency matrix as an input, in order to obtain the interconnectivity of nodes. The nodes of the input graphs can be enhanced with features to characterize their role in the graph. For our purposes, we have chosen the following features: a) the number of input and output connections to other nodes, b) the node's type (e.g., AND gate, multiplexer, etc.), and c) the number of connections to the primary inputs and outputs of the graph, if any exist. These features create vectors for each node, which are transformed into one-hot encoding representations to assist the model training.

*3) Simulation and Evaluation of FI*

At the same time, our automated flow proceeds to the evaluation of the countermeasure through FI campaigns. The adopted fault model implements a double-bit flip injection analysis, with the results serving as the ground truth used during GNN training. Our analysis first performs Single Bit-Flip (SBF) injections using Vivado simulator, for all flip-flops (FFs) of the design exhaustively (e.g., in all clock cycles and all FFs). The time of the completion of the circuit's operation is indicated by an output status signal (e.g., when the DONE signal is activated) present in the Vivado HLS-generated designs. In order to retrieve the execution time, as well as the expected response, our flow first executes a fault-free simulation with a specific test input (gold run). For the FI simulations (executed with the same specific test input), we capture the outputs of the two modules, as well as the state of the DONE signal at the completion time, defined in the gold run. We categorize the results of the FIs into the following cases: a) *"Silent"* : Correct DONE and correct output in both redundant modules; the FI has no observable effect on the outputs, b) *"Critical"*: Correct DONE and both modules have produced the same erroneous output c) *"Detected"*: Correct DONE and the outputs of the two modules are different, d)*"Hang"* : The FI has affected the DONE signal and the correctness of the output cannot be validated in either module.

After completing the SBF campaign, we perform a post-processing of the FI campaign results to deduce the Double Bit-Flip (DBF) fault model. Since the applied countermeasure is theoretically resilient to SBFs (i.e., assuming that the two replicas do not share resources, an SBF may affect only one replica), the DBF campaign is necessary for the evaluation. We examined all the combinations of two bit-flips; one bit-flip is injected into the first redundant module and one into the second module, thus excluding cases where two faults fall in the same module. This DBF evaluation does not require any additional simulations. Instead, we examine the Critical and Detected cases of each design's SBF FI campaign. Again, we take advantage of parallelization capabilities: We were able to perform multiple FI simulations at the same time, significantly speeding up the dataset generation.

*4) GNN Model Training*

This subsection describes the prediction mechanism that we developed using Python and the PyTorch Geometric library. The initial training was performed using a set of 1022 designs. Each design consists of the inputs necessary to GNNs: the adjacency matrix of each graph, the node feature vectors of the graph, transformed in one-hot encoding representations (translating to 818 input nodes for the GNN). The training label of the GNN for each design is the ground truth value resulting from the DBF FI campaigns - namely the critical error rate (CER), the detected error rate (DER), the hang error rate (HER) and the silent error rate (SER). We chose to use separate models for the training of each graph label to generate the best model for each case. To overcome training issues arising from the fact that the CERs of our target circuits range in a set of very small values (e.g., less than 1%), we used their logarithm instead of their actual value during training.

Our GNN model consists of three graph convolution layers, each followed by a ReLU activation function. After those, a global max pooling layer and a linear layer are used as the output layer. This scheme is presented in Fig. 2. As cost function during the training, we use the Mean Square Error (MSE), being the standard for regression problems. To optimize the training process, we incorporated a dynamic reduction of the learning rate, to overcome learning plateaus. If the loss function metric has stopped improving for 10 consecutive learning epochs, the mechanism lowers the learning rate parameter.

The learning rate parameter controls how a model's weights are updated in response to the loss function metric, and lowering it on plateaus results in finer training.

In order to efficiently train our GNN model, we resorted to *k*-fold cross validation. Essentially, *k* rounds of training (folds) are performed, resulting in *k* different models. For our case, *k* was empirically set to 5. Prior to the training process, we shuffled the dataset and saved 10% of the designs for a secondary evaluation (testing). The remaining 90% of the dataset was further split into 20% to be used for validation and 80% for training. In each fold, different fractions of the evaluation set were considered. Each fold was trained for a maximum of 1000 epochs, with an early stop condition set in 100 epochs after the minimum MSE loss regarding the validation set was observed. For that epoch, the fold's model was saved and re-evaluated using the test set.

There exist several graph network architectures that could be adopted in our approach regarding the convolution layer. The most prominent GNN types used in the literature are the Graph Convolutional Networks (GCN) [34] and attention-enabled graphs (GATs) [35]. We performed the training process for both cases, as well as different hyperparameters settings regarding the number of hidden nodes per convolution layer (64 and 128).

Table I presents the training results for all error rates. The results are for the Canright SBox used as a demonstration vehicle in the paper and will be discussed in the Experimental Results section. We use the R-squared achieved by the best fold model on the test dataset (which has not participated in the training or the validation process, hence is new to the model). It is calculated by summing the squared difference of predictions and ground truths, deriving it with the squared difference of ground truths to their mean value. The resulting ratio is then subtracted from 1. The R-squared value demonstrates the degree the learnable characteristics (in our case the node features) lead to the predicted graph error rate. The closer the R-squared to 1 is, the better the model predicts the actual error rates.

*B. GNN Graph Prediction Flow*

After deriving the best models for each error rate, we can use them as a prediction mechanism for the characterization of each diverse DMR design. In this case, our flow accepts as input the RTL netlist in EDIF format as depicted in Fig. 1. Then, the netlist information extraction is applied to each new design under test, followed by feature and graph adjacency matrix extraction, similar to the training flow. Instead of performing fault injections this time, the trained model can be queried to provide the prediction for the error rate of interest.

IV. DESIGN UNDER EVALUATION

For this study, we have used the high-level version of the Canright Sbox [17] as a test case. It is a well-known implementation of AES SubBytes transformation. Specifically, Canright SBox performs an inversion of a polynomial of 7th degree, with the use of the tower field approach. Furthermore, this specific implementation is readily available online and described in platform-independent C code allowing result reproduction. The Canright SBox provides an adequate design space to explore the effects of HLS directives over the functionality as well as a valid point of discussion over the need of fault resilience in cryptographic components. Module duplication was introduced as a countermeasure at the high-level C description. A wrapper function passes the 8-bit input into the two separate instances of SBox and merges their 8-bit result into a single 16-bit output. The outputs of the two SBoxes are compared externally to detect errors. The comparison logic was not included in the wrapper; the errors in the checker do not affect the reliability evaluation of the HLS designs, which is the target of our approach.

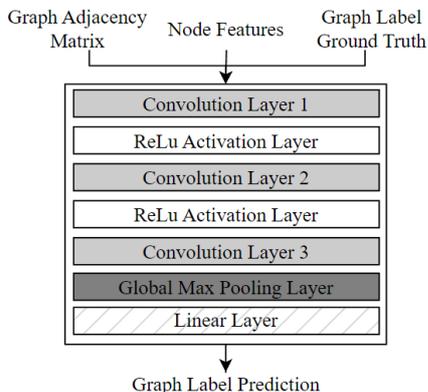

*Figure 2. The GNN Model*

*Table I. R-squared values for different settings, tested on test set*

| Model | CER | DER | HER | SER |
|---|---|---|---|---|
| CCN, 64 | 0.86 | 0.99 | 0.39 | 0.49 |
| GCN, 128 | 0.89 | 0.86 | 0.87 | 0.78 |
| GAT, 64 | 0.75 | 0.01 | 0.06 | 0.12 |
| GAT, 128 | 0.84 | -2e11 | 0.24 | 0.09 |

## V. EXPERIMENTAL RESULTS

In this section, we present and discuss the results of the training and the validation of the GNN.

### A. Ground Truth & Prediction Results Analysis

First, we examine the quality of the high-level diverse DMR by means of the fault injection campaign results. Table II presents the CER metric for three designs: a) a simple DMR implementation, b) the design in our randomly generated dataset with the lowest CER (best), and c) the design with the highest CER (worst). We focus on the CER metric since it's a key indicator for the effect of the countermeasure. Observe that since the introduction of directives was executed in a random way, the flow produced both better and worse designs than the simple DMR in terms of CER for DBF. Regarding SBFs, the DMR countermeasure presents 0% CER in all cases, depicting the successful application of DMR through HLS. For DBFs, the flow produced designs with far better CER than the simple DMR – 0.0015% CER for the best design, compared to 0.02% CER for the simple DMR application. The worst design of the dataset produced a CER of 0.3%, an order of magnitude higher than the simple DMR.

We further examined diversified DMR designs producing high CERs and noted that directives with the same or similar effect were applied to corresponding functions and loops of the two copies. That caused the elimination or reduction of diversity for those functions, worsening the CER even in comparison with the non-diversified DMR. For example, for the worst Diverse DMR case, we note the use of excessive inlining in the same code regions among the two replicas. In contrast, for the best Diverse DMR design generated by the flow, the inlining directives are mainly focused on one replica. Hence, specific directives need to be used under conditions or be entirely avoided when fault tolerance is considered. Regarding the CER model's prediction capabilities, for all cases, the difference between ground truth and prediction is less than 0.1%. Given the speedup achieved by the methodology, as shown in the following section, this error margin can be considered acceptable. Even if statistical FI is used, achieving such error margin is computationally expensive. Furthermore, the use of this prediction mechanism leaves room for improvement, by extending the examination for different models and parameterizations.

### B. GNN Error Rate Predictions

Referring to Table I, we demonstrate the successful training of the prediction mechanism on the test datasets. In all cases, the use of GCN was proved better than GAT. For the case of GCN using 128 hidden nodes among its hidden layers, all error rates approach R-squared values of 0.8, a value that empirically verifies the quality of the resulting models. The highest R-squared value among those cases was achieved by the model that predicts CER, with 0.89 and the lowest for the SER prediction, with 0.78. A notable case is the R-square value of DER achieved by a GCN model, set with 64 hidden nodes, which is equal to 0.99.

### C. Evaluation Flow Performance

Lastly, we present results demonstrating the speedup achieved by our evaluation flow. All computations were performed on an Intel® i7-9750H 6-core CPU, @2.60GHz. On average, the exhaustive SBF evaluation of each Canright SBox design took 20 minutes, while the calculation of the DBF rates took 1.5 minutes. We were able to parallelize the SBF FI campaigns by a factor of 4. A design space equal to the size of our dataset (1022) was evaluated using FI campaigns in 91.73 hours. The PyTorch Geometric Python library allowed us to perform the training process on a GPU board (NVIDIA GTX 1660), with 6GB of dedicated memory. The mean $k$-fold cross evaluation time for each model training was 2.07 hours due to the early stop condition. Up to three parallel training processes were possible at the same time, as restricted by the platform's memory size. Thus, sufficiently trained prediction models for the four error rates could be available in 95.87 hours (i.e., 91.73 hours for running the SBF experiments and calculating the DBF rates plus 2 x 2.07 hours for training the four GNN models). Afterwards, the prediction of all error rates for each design beyond the training set can be performed in ~ 4 ms.

Based on the above execution times, we can estimate the speedup achieved by our approach compared to the traditional FI campaigns and how it scales with the number of designs under evaluation. For example, the time needed to perform the necessary fault injection simulations and train the model with 1k designs takes approximately 93.73 hours. A smaller dataset (i.e. 100 designs) may not be enough to achieve the expected prediction metrics. The prediction of the error rates for a dataset of 10k designs can be performed in an additional 0.01 hours (36 seconds) if 1000 of those designs are dedicated to model training. The efficiency of our methodology becomes more prominent for larger numbers of designs; assuming the case of 1 million designs, an evaluation through FI campaigns would last 374 days, while the prediction of the error rates after the training process would only take an additional 6 minutes. The speedup achieved starts from 9.5x for 10k designs to 944.5x for 1 million designs for the given test case, as demonstrated in Table III. Lastly, it is worth noting that the design under examination is an optimistic test case regarding the evaluation methodology using FI.

*Table II. SBF & DBF CER metrics for DMR protected and Diverse DMR protected Canright SBox*

| Canright SBox Designs | SBF CER | DBF CER | Predicted DBF CER |
|---|---|---|---|
| DMR | 0 | 0.02 | 0.1 |
| Diverse DMR (best) | 0 | 0.0015 | 0.0015 |
| Diverse DMR (worst) | 0 | 0.3 | 0.21 |

*Table III. Comparison of traditional FI Campaigns with our methodology execution times (in minutes) & achieved speedup*

| #Designs | FI Campaigns | Training & Prediction | Speedup |
|---|---|---|---|
| 100 | 8.96 | - | - |
| 1,000 | 89.58 | 93.73 | - |
| 10,000 | 895.83 | 93.74 | 9.56 |
| 100,000 | 8958.33 | 93.84 | 95.47 |
| 1000000 | 89583.33 | 94.84 | 944.60 |

The examination of a more complex design (i.e. an AES hardware module) would lead to an exponential increase in the duration of its exhaustive FI examination, giving a huge advantage in adopting the proposed GNN evaluation methodology.

## VI. Conclusions & Future Work

In this work, we present an evaluation methodology of the prediction of security properties of hardware designs using state-of-the-art GNNs. Our approach is motivated by the fact that traditional fault evaluation methodologies may add significant overhead to the design and evaluation of hardware countermeasures. As a case study, we used an SBox implementation, enhanced with a diverse DMR countermeasure generated using HLS. We train the GNN model with the actual evaluation metrics extracted from exhaustive FI simulations and show that GNNs are able to take advantage of the properties of the netlist structure. Our model managed to closely approximate the values of fault metrics, regarding critical, detected, hang and silent error rates. In addition, the methodology achieved considerable speedup compared to classical FI simulation evaluations. Most importantly, the evaluation flow can be easily generalized for other designs, countermeasures as well as different fault models. We aim to continue the research over the topic, by examining more advanced GNN frameworks to further enhance the prediction quality of security metrics.

## Acknowledgment

This research has been financed by the European Union's Horizon 2020 research and innovation programme under the Marie Sklodowska-Curie grant agreement No 895937.